\documentclass[%
 reprint,
 preprintnumbers,
 amssymb,
 aps,
]{revtex4-1}

\usepackage{amsmath}
\usepackage[dvipdfmx]{graphicx}
\usepackage{dcolumn}
\usepackage{bm}
\usepackage{braket}

\newcommand{\muetoee}{$\mu^-e^-\rightarrow e^-e^-\,$}
\usepackage{color}

\definecolor{mycolor}{rgb}{0.6,0.0,0.4}

\usepackage{ulem}
\usepackage{lineno}

\begin{document}

\preprint{STUPP-17-232, J-PARC-TH-0112, MISC-2017-09}

\title{
Improved analysis for $\mu^-e^-\rightarrow e^-e^-$ in muonic atoms by photonic interaction
}

\author{Yuichi Uesaka$^1$, Yoshitaka Kuno$^1$, Joe Sato$^2$, Toru Sato$^{1,3}$, and Masato Yamanaka$^4$}
\affiliation{$^1$Department of Physics, Osaka University, Toyonaka, Osaka 560-0043, Japan\\
$^2$Physics Department, Saitama University, 255 Shimo-Okubo, Sakura-ku, Saitama, Saitama 338-8570, Japan\\
$^3$J-PARC Branch, KEK Theory Center, Institute of Particle and Nuclear Studies, KEK, Tokai, Ibaraki 319-1106, Japan\\
$^4$Maskawa Institute, Kyoto Sangyo University, Kyoto 603-8555, Japan
}




\date{\today}

\begin{abstract}
Studies of the charged lepton flavor violating process of
\muetoee in muonic atoms by the four Fermi interaction [Y. Uesaka \textit{et al}.,
Phys. Rev. D {\bf 93}, 076006 (2016)]
are extended to include the photonic interaction.
 The wave functions of a muon and electrons are obtained
 by solving the Dirac equation with the Coulomb interaction of
a finite nuclear charge distribution.
We find suppression of the \muetoee rate over the initial estimation for the photonic interaction,
in contrast to enhancement for the four Fermi interaction. It is
due to the Coulomb interaction of scattering states
and relativistic lepton wave functions.
This finding suggests that the atomic number dependence of the \muetoee
rate could be used to distinguish between the photonic
 and the four Fermi interactions.
\end{abstract}

\pacs{11.30.Hv,13.66.-a,14.60.Ef,36.10.Ee}
\maketitle

\onecolumngrid


\section{Introduction \label{sec:Introduction}}
It has been well recognized that charged lepton flavor violation (CLFV) 
is important to search for new physics beyond the standard model.
Rare processes of muons, such as
$\mu^+\rightarrow e^+\gamma$ \cite{Baldini2016},
$\mu^+\rightarrow e^+e^-e^+$ \cite{Bellgardt1988},
and $\mu^- \to e^-$ conversion \cite{Bertl2006},
have given the strongest constraints
on new physics models of CLFV interactions \cite{Kuno2001,Mori2014}. 
Furthermore, it is expected that experimental sensitivity can be significantly improved in 
near future measurements.

As a new promising process to search for CLFV interaction,
 \muetoee in a muonic atom was proposed
 \cite{Koike2010}.
For heavy atoms with large atomic numbers ($Z$), large enhancement of the \muetoee rate
due to the Coulomb attraction of the lepton wave functions to a nucleus is expected.
Another advantage for  \muetoee is that it can probe both
the four Fermi contact and the photonic interactions, as in the $\mu^+\rightarrow e^+e^-e^+$ decay and $\mu^- \to e^-$ conversion.
  In \muetoee, a sum of the energies of two electrons in the final state would be
  $m_\mu + m_e - B_\mu - B_e$, where $m_\mu$ and $m_e$ are the masses of a muon and an electron, respectively, and $B_\mu$ and $B_e$ are binding energies of the muon and electron in a muonic atom, respectively. 
The energy of each electron in the final state is about $m_\mu/2$, and they are emitted
almost back-to-back.
The search for \muetoee is proposed in
the COMET Phase-I experiment at J-PARC, Japan \cite{KEKrep}.
This new process could be essential to identify the scenario of new physics via the addition of sterile neutrinos at near future experiments \cite{Abada2016}.

The initial work \cite{Koike2010} showed that the atomic number ($Z$) dependence of the \muetoee transition rate is expected to be  of $Z^3$, owing to the
probability density of the wave functions of the Coulomb-bound electrons at origin.
This result was obtained by plane wave approximation of the outgoing
electrons and non-relativistic approximation of the bound states.
In our previous work \cite{Uesaka2016} for the case of the four Fermi contact interaction,
we have studied the Coulomb interaction for emitted electrons
by solving the Dirac equation with a finite charge distribution of nuclei.
It was found that the Coulomb interaction
is important not only for the bound leptons
but also for the electrons emitted.
 Moreover, relativistic Dirac wave functions of leptons with
 the Coulomb interaction of finite-ranged nuclear charge distribution were found to play
an important role.  As a result, the \muetoee rate increases on an atomic number $Z$
stronger than $Z^3$. For $^{208}$Pb, the \muetoee rate can be enhanced
 about 7 times larger than the previous expectation \cite{Koike2010}.
 Apparently, improved treatment of the Coulomb interaction should be
made also for the photonic contribution of \muetoee process to obtain
a complete picture of the \muetoee process.

In this work, we have made improved analyses for the \muetoee process
for the photonic interaction.
It is noticed that the photonic interaction consists of two vertices,
the  $\mu e \gamma^*$ CLFV interaction and
the  $e e \gamma^*$ electromagnetic interaction, together with
long range photon propagators.
Since an overlap integral of each vertex
involves rapidly-oscillating scattering electrons and photon wave functions
and long range Coulomb bound state wave functions,
a careful numerical study for the photonic interaction is required.
In Sec. \ref{sec:Formulation}, we start from the effective CLFV interaction
for the \muetoee process.  
The multipole expansion formula on the \muetoee rate 
is extended to the photonic interaction process.
In Sec. \ref{sec:Results}, the improved treatments of lepton
wave functions for the photonic interaction, in particular
the atomic number ($Z$) dependence of the rate, are discussed.
Then, we propose a possibility to distinguish the photonic interaction
from the four Fermi interaction, by
the atomic number ($Z$) dependence
and its angular-energy distribution of the emitted electrons.
Our analysis is summarized in Sec. \ref{sec:Conclusion}.
\section{Formulation \label{sec:Formulation}}

\subsection{Effective interaction}

The effective Lagrangian for  \muetoee consists of
the photonic interaction $\mathcal{L}_\mathrm{photo}$ and
the four Fermi interaction $\mathcal{L}_\mathrm{contact}$, as follows:
\begin{align}
  \mathcal{L}_{CLFV}=&\mathcal{L}_\mathrm{photo}+\mathcal{L}_\mathrm{contact},
\end{align}
where
\begin{align}
\mathcal{L}_\mathrm{photo}=&-\frac{4G_F}{\sqrt{2}}m_\mu
\left[A_R\overline{e_L}\sigma^{\mu\nu}\mu_R+A_L\overline{e_R}\sigma^{\mu\nu}\mu_L\right]F_{\mu\nu}+[h.c.],
\label{eq:longrange}\\
\mathcal{L}_\mathrm{contact}=&-\frac{4G_F}{\sqrt{2}}
[g_1(\overline{e_L}\mu_R)(\overline{e_L}e_R)
+g_2(\overline{e_R}\mu_L)(\overline{e_R}e_L) \nonumber\\
&+g_3(\overline{e_R}\gamma_\mu\mu_R)(\overline{e_R}\gamma^\mu e_R)
+g_4(\overline{e_L}\gamma_\mu\mu_L)(\overline{e_L}\gamma^\mu e_L) \nonumber\\
&+g_5(\overline{e_R}\gamma_\mu\mu_R)(\overline{e_L}\gamma^\mu e_L)
+g_6(\overline{e_L}\gamma_\mu\mu_L)(\overline{e_R}\gamma^\mu e_R)]+[h.c.].
\label{eq:shortrange}
\end{align}
Here, $G_F=1.166\times10^{-5}$GeV$^{-2}$ is the Fermi coupling constant, and
$A_{L/R}$ and $g_i$ ($i=1,\cdots,6$) are the coupling constants which are determined by new physics models.
The left- and right-handed fields $\psi_{L/R}$ are defined as $\psi_{L/R}=P_{L/R}\psi$, using the projection operators $P_{L/R}=(1\mp\gamma_5)/2$.

\begin{figure}[htbp]
  \centering
    \begin{tabular}{c}
      \begin{minipage}{0.49\hsize}
		\centering
          \includegraphics[clip, width=4.5cm]{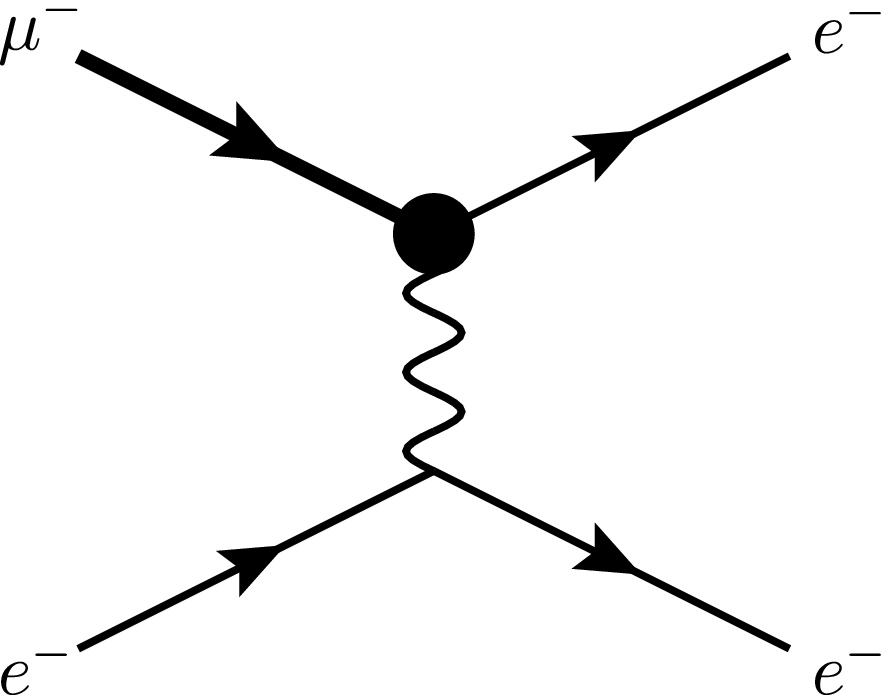}
          \\ (a)
      \end{minipage}%
      \begin{minipage}{0.49\hsize}
        \centering
          \includegraphics[clip, width=4.5cm]{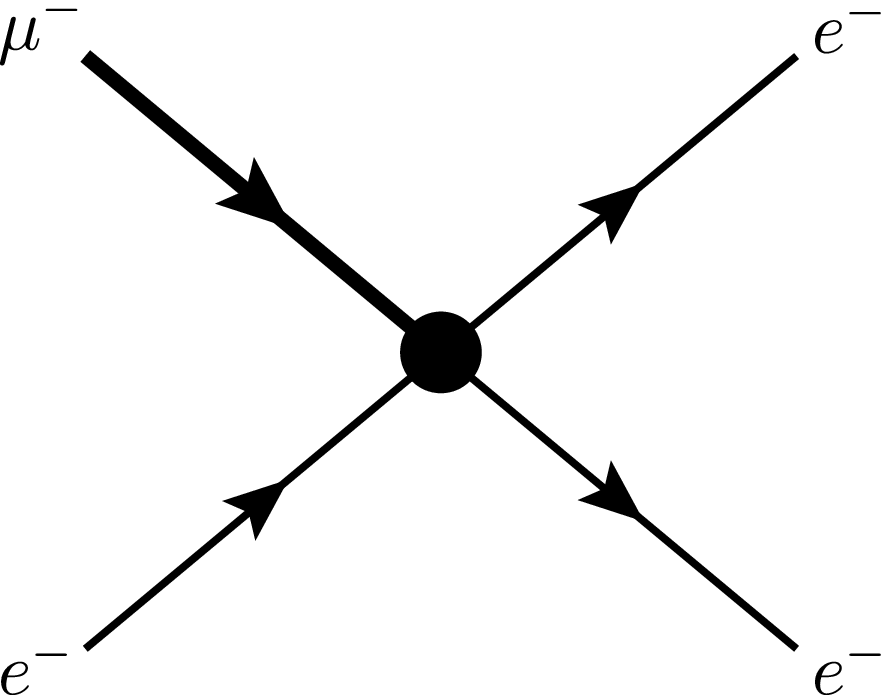}
          \\ (b)
      \end{minipage}%
    \end{tabular}
    \caption{The diagrams representing \muetoee: the one-photon-exchange
      photonic interaction (a) and the four Fermi contact interaction (b). The black closed circle shows the CLFV interaction.}
    \label{fig:diagram}
\end{figure}
The one-photon-exchange photonic interaction shown in Fig.~\ref{fig:diagram} (a)
is given by the photonic interaction in Eq. (\ref{eq:longrange}) together with
the electromagnetic interaction of
\begin{align}
\mathcal{L}_{em}= -q_e \overline{e}\gamma^\lambda eA_\lambda.
\end{align}
Here $q_e=-e$ is a charge of an electron.
The four Fermi interaction shown in Eq. (\ref{eq:shortrange})
and Fig.~\ref{fig:diagram} (b) has been studied \cite{Uesaka2016}.
The transition amplitude $M$ of \muetoee is given by,
\begin{eqnarray}
  2\pi i \delta(E_\mathrm{f}-E_\mathrm{i})
  M(\bm{p}_1,s_1,\bm{p}_2,s_2;\alpha_\mu,s_\mu,\alpha_e,s_e)
  & = & 
  \braket{e^{s_1}_{\bm{p}_1}e^{s_2}_{\bm{p}_2}|T[\exp\left\{i\int d^4x
       (\mathcal{L}_{CLFV} + \mathcal{L}_{em}) 
      \right\}]
    |\mu^{s_\mu}_{1S}e^{s_e}_{\alpha_e}},
\end{eqnarray}
with
\begin{align}
  M(\bm{p}_1,s_1,\bm{p}_2,s_2;\alpha_\mu,s_\mu,\alpha_e,s_e)=&
  M_\mathrm{photo}(\bm{p}_1,s_1,\bm{p}_2,s_2;\alpha_\mu,s_\mu,\alpha_e,s_e)+
  M_\mathrm{contact}(\bm{p}_1,s_1,\bm{p}_2,s_2;\alpha_\mu,s_\mu,\alpha_e,s_e).
\label{eq:M_total}
\end{align}
Here $E_i$ and $E_f$ are the energy of the initial and final state
given as $E_i= m_\mu-B_\mu^{1S}+m_e-B_e^{\alpha_e}$ and $E_f= E_{p_1}+E_{p_2}$, respectively.
And $E_{p_i}$ is an energy of the electron with its momentum $p_i$
and $B_l^\alpha$ is a binding energy of the lepton $l$ in the state $\alpha$.
The principle quantum number $n$ and $\kappa$
\cite{rose1961,rose1995elementary} of the bound muon and electron are
collectively denoted by $\alpha_\mu$ and $\alpha_e$, respectively.
We assume the initial muon is in its $1S_{1/2}$ ($n=1$ and $\kappa=-1$) state,
while we have included contribution of all bound electrons.
The expression of $M_\mathrm{contact}$ is given as $M$ in Eq. (4) of Ref.~\cite{Uesaka2016}.
The amplitude of the photonic interaction $M_\mathrm{photo}$ is given as
\begin{align}
  M_\mathrm{photo}(\bm{p}_1,s_1,\bm{p}_2,s_2;1S,s_\mu,\alpha_e,s_e)=
  &\left[\frac{8G_F}{\sqrt{2}}m_\mu q_e\int d^3x_1d^3x_2G_\nu\left(\bm{x}_1,\bm{x}_2;m_\mu-B_\mu^{1S}-E_{p_1}\right)\right. \nonumber\\
&\left.\times\overline{\psi}_{\bm{p}_1,s_1}^e(\bm{x}_1)\sigma^{\mu\nu}\left(A_LP_L+A_RP_R\right)\psi_{1S,s_\mu}^{\mu}(\bm{x}_1)\overline{\psi}_{\bm{p}_2,s_2}^e(\bm{x}_2)\gamma_\mu\psi_{\alpha_e,s_e}^{e}(\bm{x}_2)\right] \nonumber\\
&-\left[\left\{\bm{p}_1,s_1\right\}\leftrightarrow\left\{\bm{p}_2,s_2\right\}\right].
\label{eq:mphoton}
\end{align}
The second term $\left\{\bm{p}_1,s_1\right\}\leftrightarrow\left\{\bm{p}_2,s_2\right\}$ is 
obtained by exchanging the quantum numbers of the final electrons in the first term.
The photonic interaction is a finite range interaction between the two leptons and 
$G_\nu\left(\bm{x}_1,\bm{x}_2;q_0\right)$ is defined  as
\begin{align}
G_\nu\left(\bm{x}_1,\bm{x}_2;q_0\right)=&\int\frac{d^3q}{(2\pi)^3}\frac{iq_\nu e^{-i\bm{q}\cdot\left(\bm{x}_1-\bm{x}_2\right)}}{|\bm{q}|^2-q_0^2-i\epsilon}.
\end{align}

\subsection{Multipole expansion}

To proceed, we derive a multipole expansion of the transition amplitude.
Based on a standard partial wave expansion of the scattering wave functions
and the bound state wave functions of Dirac particles
given in Eqs. (11), (12), and (13) of Ref.~\cite{Uesaka2016},
the transition amplitude is expressed as
\begin{align}
M(\bm{p}_1,s_1,\bm{p}_2,s_2;1S,s_\mu,\alpha_e,s_e)=& 2\sqrt{2}G_F\sum_{\kappa_1,\kappa_2,\nu_1,\nu_2,m_1,m_2}\left(4\pi\right)^2\left(-i\right)^{l_{\kappa_1}+l_{\kappa_2}}e^{i\left(\delta_{\kappa_1}+\delta_{\kappa_2}\right)} \nonumber\\
&\times Y_{l_{\kappa_1},m_1}\left(\hat{p}_1\right)Y_{l_{\kappa_2},m_2}\left(\hat{p}_2\right)\left(l_{\kappa_1},m_1,1/2,s_1|j_{\kappa_1},\nu_1\right)\left(l_{\kappa_2},m_2,1/2,s_2|j_{\kappa_2},\nu_2\right) \nonumber\\
&\times\sum_{J,M}\left(j_{\kappa_1},\nu_1,j_{\kappa_2},\nu_2|J,M\right)\left(j_{-1},s_\mu,j_{\kappa_e},s_e|J,M\right) \nonumber\\
&\times\frac{\sqrt{2\left(2j_{\kappa_1}+1\right)\left(2j_{\kappa_2}+1\right)\left(2j_{\kappa_e}+1\right)}}{4\pi}N\left(J,\kappa_1,\kappa_2,E_{p_1},\alpha_e\right),
\label{eq:M_photo}
\end{align}
where $\left(l_\kappa,m,1/2,s|j_\kappa,\nu\right)$ and $Y_{l_\kappa,m}(\hat{p})$ are the Clebsch-Gordan coefficients and the spherical harmonics, respectively.
Here $l_\kappa,j_\kappa$ are the orbital and the total angular momentum of the state with $\kappa$.
$\delta_\kappa$ is a phase shift of the scattering state.
The partial wave amplitude, $N\left(J,\kappa_1,\kappa_2,E_{p_1},\alpha_e\right)$ for the photonic  and the contact interactions is given by
\begin{align}
  N\left(J,\kappa_1,\kappa_2,E_{p_1},\alpha_e\right)= N_\mathrm{photo} + N_\mathrm{contact},
\end{align}
with
\begin{eqnarray}
  N_\mathrm{photo} &=&\sum_{i=L/R}A_iW_i(J,\kappa_1,\kappa_2,E_{p_1},\alpha_e)\\
  N_\mathrm{contact}& = &\sum_{i=1}^{6}g_iW_i(J,\kappa_1,\kappa_2,E_{p_1},\alpha_e).
\end{eqnarray}
Here $W_i$s ($i=1,2,\cdots,6$) for the contact interaction are given in Ref.~\cite{Uesaka2016}.
The amplitudes of the photonic interaction $W_{L/R}$ are given as
\begin{align}
W_{L/R}=&\frac{2m_\mu}{i}\sqrt{\pi\alpha}\sum_{l=0}^{\infty}\sum_{j=|l-1|}^{l+1}\sum_{\lambda=1}^{3}\left[X_\lambda\left(l,j,\kappa_1,\kappa_2,J\right)\pm iY_\lambda\left(l,j,\kappa_1,\kappa_2,J\right)\right],
\label{eq:W_LR}
\end{align}
where $\pm$ corresponds to $L$ and $R$, respectively.
$X_\lambda$ and $Y_\lambda$ are expressed in terms of $Z$ as
\begin{align}
X_1\left(l,j,\kappa_1,\kappa_2,J\right)=&(-1)^{l+j}\left\{Z_{gfgf}^{l,l,1,j}(J)+Z_{fggf}^{l,l,1,j}(J)-Z_{gffg}^{l,l,1,j}(J)-Z_{fgfg}^{l,l,1,j}(J)\right\}, \\
X_2\left(l,j,\kappa_1,\kappa_2,J\right)=& f_{l-j}^{(2)}(j)\left\{Z_{gfgg}^{l,j,0,j}(J)+Z_{fggg}^{l,j,0,j}(J)+Z_{gfff}^{l,j,0,j}(J)+Z_{fgff}^{l,j,0,j}(J)\right\}, \\
X_3\left(l,j,\kappa_1,\kappa_2,J\right)=& f_{l-j}^{(3)}(j)\sum_{\{l_a,l_b\}=\{l,j\},\{j,l\}}\left\{Z_{gggf}^{l_a,l_b,1,j}(J)-Z_{ffgf}^{l_a,l_b,1,j}(J)-Z_{ggfg}^{l_a,l_b,1,j}(J)+Z_{fffg}^{l_a,l_b,1,j}(J)\right\},
\end{align}
\begin{align}
Y_1\left(l,j,\kappa_1,\kappa_2,J\right)=&(-1)^{l+j}\left\{Z_{gggf}^{l,l,1,j}(J)-Z_{ffgf}^{l,l,1,j}(J)-Z_{ggfg}^{l,l,1,j}(J)+Z_{fffg}^{l,l,1,j}(J)\right\}, \\
Y_2\left(l,j,\kappa_1,\kappa_2,J\right)=& f_{l-j}^{(2)}(j)\left\{Z_{gggg}^{l,j,0,j}(J)-Z_{ffgg}^{l,j,0,j}(J)+Z_{ggff}^{l,j,0,j}(J)-Z_{ffff}^{l,j,0,j}(J)\right\}, \\
Y_3\left(l,j,\kappa_1,\kappa_2,J\right)=& f_{l-j}^{(3)}(j)\sum_{\{l_a,l_b\}=\{l,j\},\{j,l\}}\left\{Z_{gffg}^{l_a,l_b,1,j}(J)+Z_{fgfg}^{l_a,l_b,1,j}(J)-Z_{gfgf}^{l_a,l_b,1,j}(J)-Z_{fggf}^{l_a,l_b,1,j}(J)\right\},
\end{align}
where
\begin{align}
f_{h}^{(2)}(j)=
\begin{cases}
\sqrt{\dfrac{j+1}{2j+1}} & (h=+1) \\
0 & (h=0) \\
\sqrt{\dfrac{j}{2j+1}} & (h=-1)
\end{cases}, \hspace{10mm}
f_{h}^{(3)}(j)=
\begin{cases}
\sqrt{\dfrac{j}{2j+1}} & (h=+1) \\
0 & (h=0) \\
-\sqrt{\dfrac{j+1}{2j+1}} & (h=-1)
\end{cases}.
\end{align}

The matrix element $Z$, which consists of CLFV and the electromagnetic vertex and
the photon propagator is given by,
\begin{align}
  Z_{ABCD}^{l_a,l_b,s,j}(J)\equiv&\left[q_0^2
    \int_{0}^{\infty}dr_1r_1^2A_{p_1}^{\kappa_1}(r_1)B_{1,\mu}^{\kappa_\mu}(r_1)
    \int_{0}^{\infty}dr_2r_2^2F_{l_a,l_b}^{q_0}(r_1,r_2)C_{p_2}^{\kappa_2}(r_2)
    D_{n,e}^{\kappa_e}(r_2)\right. \nonumber\\
&\left.\times(-1)^{J+\kappa_2+\kappa_e}
V_{l_a,1,j}^{s_A\kappa_1,s_B\kappa_\mu}V_{l_b,s,j}^{s_C\kappa_2,s_D\kappa_e}
W\left(j_{\kappa_1}j_{\kappa_2}1/2j_{\kappa_e};Jj\right)\right] \nonumber\\
&-(-1)^{j_{\kappa_1}+j_{\kappa_2}-J}\left[\left\{p_1,\kappa_1\right\}\leftrightarrow\left\{p_2,\kappa_2\right\}\right],
\label{eq:Z_ABCD}
\end{align}
where $\kappa_\mu = -1$ and $W(abcd;ef)$ is the Racah coefficient.
Here,
  $A_{p}^{\kappa}(r), C_{p}^\kappa(r)$ and $B_{n,\mu}^{\kappa}(r), D_{n,e}^{\kappa}(r)$
  are radial wave functions of the scattering states ($g_p^\kappa, f_p^\kappa$)
  and the bound states ($g_{n,l}^\kappa, f_{n,l}^\kappa$) given in Appendix
  and $s_A = \pm 1$ for $A=g$ and $A=f$, respectively.
 $q_0$ is $q_0=m_\mu-B_\mu^{1S}-E_{p_1}$ for the direct term,
and $q_0=m_\mu-B_\mu^{1S}-E_{p_2}$ for the exchange term.
The partial wave expansion of the photon propagator is given as
\begin{align}
  \int\frac{d^3q}{(2\pi)^3}\frac{q_\nu e^{-i\bm{q}\cdot\left(\bm{x}_1-\bm{x}_2\right)}}{|\bm{q}|^2-q_0^2-i\epsilon}=q_0\partial_\nu\sum_{l,m}Y_{l,m}^*\left(\hat{x}_1\right)Y_{l,m}\left(\hat{x}_2\right)F_{l,l}^{q_0}\left(x_1,x_2\right),
  \label{eq:photong}
\end{align}
where we have defined $\partial_{\nu}=(iq_0,\bm{\nabla}_1)$ and
\begin{align}
F_{l_1,l_2}^{q_0}\left(x_1,x_2\right)=h_{l_1}^{(1)}\left(q_0x_1\right)j_{l_2}\left(q_0x_2\right)\theta(x_1-x_2)+h_{l_2}^{(1)}\left(q_0x_2\right)j_{l_1}\left(q_0x_1\right)\theta(x_2-x_1).
\end{align}
Here $j_l$ and $h_l^{(1)}$ are the spherical Bessel function and the first kind
spherical Hankel function, respectively.
The radial integral of the CLFV vertex is extended in a range of the Bohr radius of the muon,
while the integrand extends to the electron Bohr radius for the electromagnetic vertex.
Since the wave length of the electron scattering state around 50MeV is about 1/4 fm,
a numerical integration for this radial integral is carefully made.
The coefficient $V$s are given by reduced matrix elements of the spin-orbital
wave function by;
\begin{align}
V_{l,s,j}^{\kappa_b,\kappa_a}=&(-1)^{l}\frac{1+(-1)^{l_{\kappa_b}+l_{\kappa_a}+l}}{2}\left(j_{\kappa_b},1/2,j_{\kappa_a},-1/2|j,0\right) \nonumber\\
&\times
\begin{cases}
\delta_{l,j} & (s=0, j=l) \\
(j-\kappa_a-\kappa_b)/\sqrt{j(2j+1)} & (s=1, j=l+1) \\
(\kappa_a-\kappa_b)/\sqrt{j(j+1)} & (s=1, j=l) \\
-(j+1+\kappa_a+\kappa_b)/\sqrt{(j+1)(2j+1)} & (s=1, j=l-1)
\end{cases}.
\end{align}

Finally, the angular and energy distributions of the emitted electron are expressed
in terms of the partial wave amplitude by
\begin{align}
\frac{d^2\Gamma_{\alpha_e}}{dE_{p_1}d\cos\theta}=&\frac{G_F^2}{2\pi^3}|\bm{p}_1||\bm{p}_2|\sum_{\kappa_1,\kappa_2,\kappa'_1,\kappa'_2,J,l}(2J+1)(2j_{\kappa_e}+1)\left(2j_{\kappa_1}+1\right)\left(2j_{\kappa_2}+1\right)\left(2j_{\kappa_1'}+1\right)\left(2j_{\kappa_2'}+1\right) \nonumber\\
&\times\frac{1+(-1)^{l_{\kappa_1}+l_{\kappa'_1}+l}}{2}\frac{1+(-1)^{l_{\kappa_2}+l_{\kappa'_2}+l}}{2}i^{-l_{\kappa_1}-l_{\kappa_2}+l_{\kappa_1'}+l_{\kappa_2'}}e^{i\left(\delta_{\kappa_1}+\delta_{\kappa_2}-\delta_{\kappa_1'}-\delta_{\kappa_2'}\right)} \nonumber\\
&\times(j_{\kappa_1},1/2,j_{\kappa'_1},-1/2|l,0)(j_{\kappa_2},1/2,j_{\kappa'_2},-1/2|l,0)W(j_{\kappa_1}j_{\kappa_2}j_{\kappa'_1}j_{\kappa'_2};Jl) \nonumber\\
&\times(-1)^{J-j_{\kappa_2}-j_{\kappa'_2}}N(J,\kappa_1,\kappa_2,E_{p_1},\alpha_e)N^*(J,\kappa'_1,\kappa'_2,E_{p_1},\alpha_e)P_l(\cos\theta),
\end{align}
where $P_l(x)$ is Legendre polynomials.
The total rate can be calculated by integrating the
energy and angle:
\begin{align}
\Gamma=\frac{1}{2}\sum_{\alpha_e}\int_{m_e}^{m_\mu-B_\mu^{1S}-B_e^{\alpha_e}}dE_{p_1}\int_{-1}^{1}d\cos\theta\frac{d^2\Gamma_{\alpha_e}}{dE_{p_1}d\cos\theta}.
\label{gamma-new}
\end{align}
After taking into account the approximations employed in  Ref.~\cite{Koike2010},
the above formula for the photonic interaction can be reduced to the following transparent formula of, 
\begin{align}
\Gamma_0(\mu^-e^-\rightarrow e^-e^-)
 =\frac{8m_e}{\pi}(Z-1)^3\alpha^4\left(G_Fm_\mu^2\right)^2\left(\left|A_R\right|^2+\left|A_L\right|^2\right).
\label{eq:Gamma0_photo}
\end{align}


\section{Results \label{sec:Results}}

The wave functions of the bound muon and electron and the emitted electrons in the final state
are obtained by solving Dirac equations with the Coulomb potential numerically.
We use the uniform nuclear charge distribution, $\rho_C(r)$, for the
Coulomb potential, which is given as
\begin{align}
\rho_C(r)=\frac{3Ze}{4\pi R^3}\theta(R-r),
\end{align}
with $R=1.2A^{1/3}$fm. We have also examined
a realistic charge distribution of the Woods-Saxon form. However the
rate changes by less than 1\% from that of the uniform distribution.
Therefore the uniform charge distribution is decided to use in our calculation from now on.
A sufficiently large number of partial waves of the scattering electron state has
to be included. The convergence property of the rate against partial waves
is shown in Table \ref{tab:convergence}. 
\begin{table}[htb]
\setlength{\extrarowheight}{3pt}
\centering
\caption{The convergence property of $\Gamma/\Gamma_0$.
The maximum values of $|\kappa|$ included for the rates in each column are given in the first low.}
¡¡\vspace{0.4cm}
¡¡\doublerulesep 0.8pt \tabcolsep 0.4cm
  \begin{tabular}{ccccc} \hline\hline
    Nuclei & $|\kappa|\le1$ & $|\kappa|\le5$ & $|\kappa|\le10$ & $|\kappa|\le20$ \\ \hline
    $^{40}$Ca & 0.0762 & 0.482 & 0.641 & 0.663 \\
    $^{120}$Sn & 0.125 & 0.396 & 0.406 & 0.406 \\
    $^{208}$Pb & 0.109 & 0.270 & 0.271 & 0.271 \\ \hline\hline
  \end{tabular}
  \label{tab:convergence}
\end{table}
The convergence property  is almost the same as the
contact interaction. For $^{40}$Ca, we have to sum the partial waves up
to $|\kappa|\le20$. For larger $Z$ nuclei, the rate converges
faster due to a smaller radius of the bound muon.

For the photonic interaction, the effective interaction is not local
due to the propagation of virtual photons. In principle, the bound electrons
other than 1S state could contribute to the rate.
The contributions to the rate from each atomic orbit, normalized to those of 1S state,
are shown in Table \ref{tab:higher_shells}.
It is found the contributions of the non-S wave bound electrons are larger than
that of the contact interaction. However it is still very small, compared with those of
1S electrons. The total rate for $^{208}$Pb is enhanced by about 20\% by including
the electrons other than the 1S state.
\begin{table}[htb]
\setlength{\extrarowheight}{3pt}
\centering
\caption{The relative contributions to the rate from electrons in different atomic orbits under
  the $M$ shell and the $4S$ orbit, for $^{208}$Pb. The spins are summed over. They are normalized by the 1S
  contribution.}
¡¡\vspace{0.4cm}
¡¡\doublerulesep 0.8pt \tabcolsep 0.4cm
  \begin{tabular}{cccccccc} \hline\hline
    1S & $2S$ & $2P$ & $3S$ & $3P$ & $3D$ & $4S$ & Total \\ \hline
    1 & 0.15 & 7.3$\times 10^{-3}$ & 4.3$\times 10^{-2}$ & 2.6$\times 10^{-3}$ & 2.5$\times 10^{-5}$ & 1.8$\times 10^{-2}$ & 1.21 \\ \hline\hline
  \end{tabular}
  \label{tab:higher_shells}
\end{table}

\subsection{Rate of the photonic interaction}

\begin{figure}[htb]
	\centering
	\includegraphics[width=70mm]{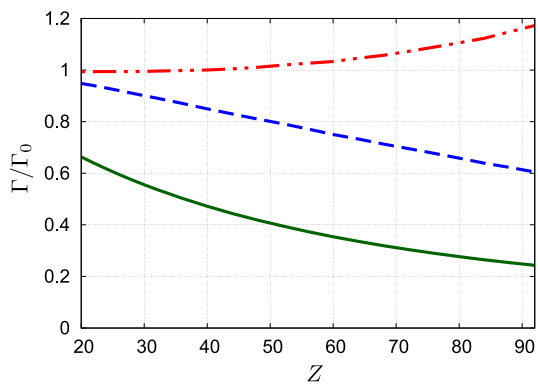}
	\caption{The $Z$ dependence of $\Gamma/\Gamma_0$. 
        The ratios $\Gamma/\Gamma_0$ of model I, II and III
        are shown in dashed, dash-two-dotted, and solid curves.}
	\label{fig:decayrate_AL}
\end{figure}

The ratio of the \muetoee rates, $\Gamma/\Gamma_0$, is studied
to examine the roles of Coulomb interaction of the scattering state
and the relativistic wave function of the bound states.
For simplicity, we set $A_R=0$ and start discussions including
 only the contribution of the 1S electron bound state.
We introduce three models summarized in Table \ref{tab:model}.
In the model I, the scattering electrons are plane waves (PLW) 
and the wave function of bound electron is non-relativistic (Non. Rel.).
The ratio of $\Gamma/\Gamma_0$ is shown in a dashed line of Fig.~\ref{fig:decayrate_AL}.
Due to the finite size of the muon wave function, it is decreasing linearly as
$Z$, even though the approximation for the lepton wave functions are
the same in this work and Ref.~\cite{Koike2010}.
This is observed in our previous work \cite{Uesaka2016} for the contact interaction.
In the model II, we replaced the bound state wave functions in the model I 
by relativistic one (Rel.).
The result is shown in a dash-two-dotted line.
The relativistic effect of a bound electron at small distance
makes the overlap integral larger and the ratio becomes $1 \sim 1.2$.
Finally, we use the Coulomb distorted wave (DW) for the electron scattering state
in the model III. The ratio is shown in a solid line.
By taking into account the Coulomb distortion and the relativistic bound state
wave functions, the rate is strongly suppressed compared with $\Gamma_0$,
which is quite different from large enhancement obtained for the contact interaction.
The ratio is 0.27 (0.66) for $^{208}$Pb ($^{40}$Ca).

\begin{table}[htb]
\setlength{\extrarowheight}{3pt}
\centering
¡¡\caption{Models for the electron wave functions. 
The relativistic bound state wave function (Rel.) and distorted wave of the scattering state (DW) 
are calculated in by the Coulomb potential from an uniform nuclear charge density.
The non-relativistic bound state wave function (Non.) is obtained by using a point charge density.}
¡¡\vspace{0.4cm}
¡¡\doublerulesep 0.8pt \tabcolsep 0.4cm
  \begin{tabular}{ccc} \hline\hline
  Model & Bound electron & Scattering electron \\ \hline
  I     & Non. Rel. & PLW \\
  II     & Rel. & PLW \\
  III     & Rel. & DW \\  \hline\hline
  \end{tabular}
  \label{tab:model}
\end{table}

\begin{figure}[htb]
	\centering
	\includegraphics[width=70mm]{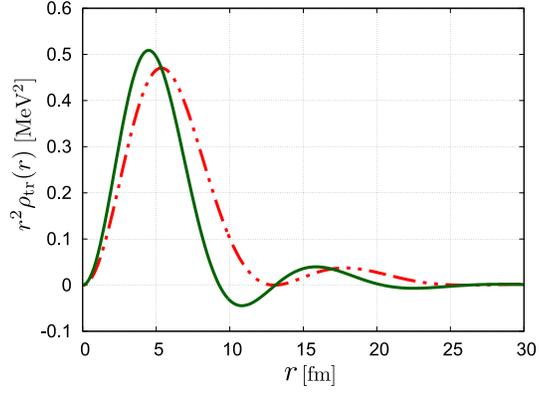}
	\caption{
	The transition density $r^2\rho_\mathrm{tr}(r)$ for $^{208}$Pb.
	The dash-two-dotted and solid curves show the transition density using PLW and DW scattering electron, respectively.
	Here, the bound muon is treated relativistically in both curves.
	}
	\label{fig:overlap_mu}
\end{figure}

To understand the mechanism of the suppression of the \muetoee rate, we study a typical transition density,
\begin{align}
\rho_\mathrm{tr}(r)=j_0\left(q_0r\right)g_{p_1}^{-1}(r)g_{1,\mu}^{-1}(r),
\end{align}
which indicates the partial transition density of a bound muon (1S) to a scattering electron ($\kappa=-1$) and a photon ($l=0$).
Here we select the most important kinematical region $p_1 = (m_\mu - B_\mu)/2 = q_0$, ignoring the electron mass.
The transition densities calculated by using the PLW and DW electron wave functions
are shown in Fig.~\ref{fig:overlap_mu}.
In the PLW case, $\rho_\mathrm{tr}$ is positive definite, since the wave length of the scattering electrons is the same as that of virtual photons.
On the other hand, $\rho_\mathrm{tr}$ changes its sign and oscillates because of the Coulomb attraction for the electron.
The same mechanism also can be applied to the vertex of the bound electron transition.
Therefore the distortion of final electrons suppresses the transition rate.

In terms of the momentum space,
the suppression of the \muetoee rate for the photonic interaction can be understood as follows.
The momenta of  the electron and virtual photon are transferred to the bound muon or the electron at each vertex for the photonic interaction.
Main contribution to the rate is 
when the both electron and virtual photons carry about a half of the muon mass,
so that the momentum transfer to the bound states is almost zero.
While this is true for the asymptotic momentum of the electron,
the Coulomb attraction increases local momentum of the electrons being close to the nucleus.
This brings a mismatch of the
virtual photon and the electron momenta and increases the momentum transfer to the bound leptons
and hence the transition probability is reduced.
A similar suppression mechanism of the transition rate
was pointed out in Ref.~\cite{Shanker1979} for the $\mu^-\to e^-$ conversion process.

\begin{figure}[htb]
\centering
\includegraphics[width=70mm]{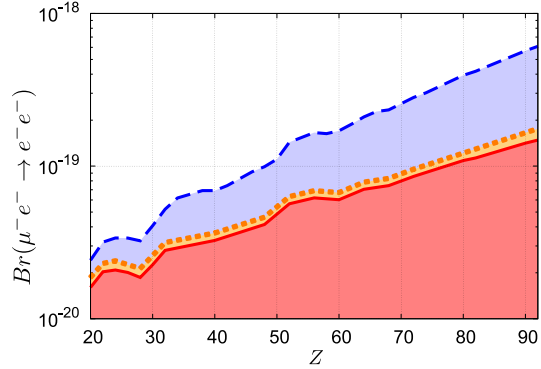}
\caption{Upper limits on $Br(\mu^-e^-\rightarrow e^-e^-)$,
  constrained by the experimental upper limits of
  $Br(\mu^+\rightarrow e^+\gamma)<4.2\times 10^{-13}$ \cite{Baldini2016}.
The dashed (blue) curve shows the result of previous work \cite{Koike2010}.
Our results including only the 1S electrons and all the 1S electrons are shown by the solid (red) and the dotted (orange) lines, respectively.}
\label{fig:branchingratio_AL}
\end{figure}

The branching ratio of \muetoee by the photonic interaction is given as
\begin{align}
Br(\mu^-e^-\rightarrow e^-e^-)\equiv\tilde{\tau}_\mu\Gamma(\mu^-e^-\rightarrow e^-e^-),
\end{align}
where $\tilde{\tau}_\mu$ is a mean life time of the  muonic atom, given in Ref.~\cite{Suzuki1987}.
  The upper limit of this branching ratio
  is calculated by using  $A_R$ and $A_L$, which are
  constrained from the  experimental upper limit 
  of $\mu^+\rightarrow e^+\gamma$.
The branching ratio
$Br(\mu^+\rightarrow e^+\gamma) = \Gamma(\mu^+\rightarrow e^+\gamma)/\Gamma(\mu^+ \rightarrow e^+\overline{\nu}_\mu \nu_e)$
is given as
\begin{align}
Br(\mu^+\rightarrow e^+\gamma)=384\pi^2\left(|A_R|^2+|A_L|^2\right).
\end{align}
Assuming the dominance of the photonic interaction, 
the upper limit of $Br(\mu^-e^-\rightarrow e^-e^-)$ can be expressed
  by using $B_\mathrm{max}$, which is current upper limit of $Br(\mu^+\rightarrow e^+\gamma)$
 as,
\begin{align}
Br(\mu^-e^-\rightarrow e^-e^-)<&\frac{Br(\mu^-e^-\rightarrow e^-e^-)}{Br(\mu^+\rightarrow e^+\gamma)}B_\mathrm{max} \nonumber\\
=& 4(Z-1)^3\alpha^4\frac{m_e}{m_\mu}\frac{\tilde{\tau}_\mu}{\tau_\mu}
    \frac{\Gamma(\mu^-e^-\rightarrow e^-e^-)}{\Gamma_0(\mu^-e^-\rightarrow e^-e^-)}
       B_\mathrm{max},
\label{eq:brmax}
\end{align}
where $\tau_\mu$ is the mean life time of a free muon.
  The upper limit of the branching ratio (Eq. (\ref{eq:brmax}))
  is calculated as a function of $Z$ by using $B_\mathrm{max}=4.2\times 10^{-13}$ by the MEG experiment \cite{Baldini2016}.
  The dashed (blue) line in Fig.~\ref{fig:branchingratio_AL}
  shows the result of previous work \cite{Koike2010}, whereas 
  the results of this work with taking into account the 1S electrons and all the bound electrons are shown in a solid (red) and dotted (orange) lines, respectively.
  From the improved estimations using the relativistic Coulomb lepton wave functions,
  the branching ratio $Br(\mu^-e^-\rightarrow e^-e^-)$ is about $10^{-19}$ for $^{208}$Pb.
  The non-1S bound electrons increase the branching ratio by about $20$\%.

\subsection{Distinguishing mechanisms of CLFV interactions}

\begin{figure}[htb]
	\centering
	\includegraphics[width=70mm]{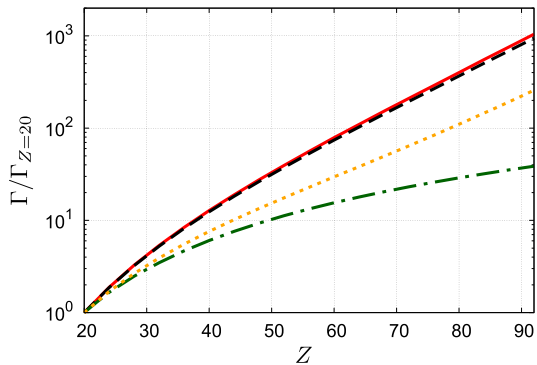}
	\caption{$Z$ dependence of \muetoee generated by four different models. They are normalized by the rate for $Z=20$. A solid red line shows the case of model (i), a dashed black line shows that of model (ii), a dash-dotted green one shows that of model (iii), and a dotted orange one shows that of model (iv).}
	\label{fig:decayrate_g1_g5_AL}
\end{figure}

Having completed to study
the \muetoee process
for both the contact and the photonic interactions,
we study a possibility to distinguish the CLFV mechanism of the
\muetoee process in muonic atoms.
For this purpose we consider four simplified models:
(i) contact interaction, where the electrons are emitted with the same chirality.
\begin{align}
g_1\ne 0, \hspace{3mm} A_{L/R}=0, \hspace{3mm}\mbox{and} \ \  g_{j \neq 1}=0,
\end{align}
(ii) contact interaction, where the electrons are emitted with opposite chirality.
\begin{align}
g_5\ne 0, \hspace{3mm} A_{L/R}=0, \hspace{3mm}\mbox{and} \ \  g_{j \neq 5}=0,
\end{align}
(iii) photonic interaction
\begin{align}
A_L\ne 0, \hspace{3mm} A_R=0, \hspace{3mm}\mbox{and} \ \  g_i=0.
\end{align}
(iv) both of contact and photonic interactions
\begin{align}
g_1=100A_L\ne 0, \hspace{3mm} A_R=0, \hspace{3mm}\mbox{and} \ \  g_{j \neq 1}=0.
\end{align}
We have chosen $g_1/A_L=100$ in the model (iv), while $g_1/A_L\sim 270$ using the current upper limits of $A_L$ and $g_1$.
The $Z$ dependence of \muetoee  is shown in Fig.~\ref{fig:decayrate_g1_g5_AL}.
The ratios of the models (i) (in a solid line) and (ii) (in a dashed line) strongly increase as $Z$.
One would need precise measurements to discriminate the model (i) from (ii).
On the other hand, the model (iii) exhibits a moderately increase as $Z$.
We may expect the contribution from both the photonic and the contact interactions in the model (iv) and the $Z$ dependence is drawn as a dotted line in Fig.~\ref{fig:decayrate_g1_g5_AL}.
Thus, we can distinguish the CLFV interactions and their dominance by the $Z$ dependence of \muetoee.

\begin{figure}[htb]
  \centering
    \begin{tabular}{c}
      \begin{minipage}{0.49\hsize}
		\centering
		\includegraphics[width=70mm]{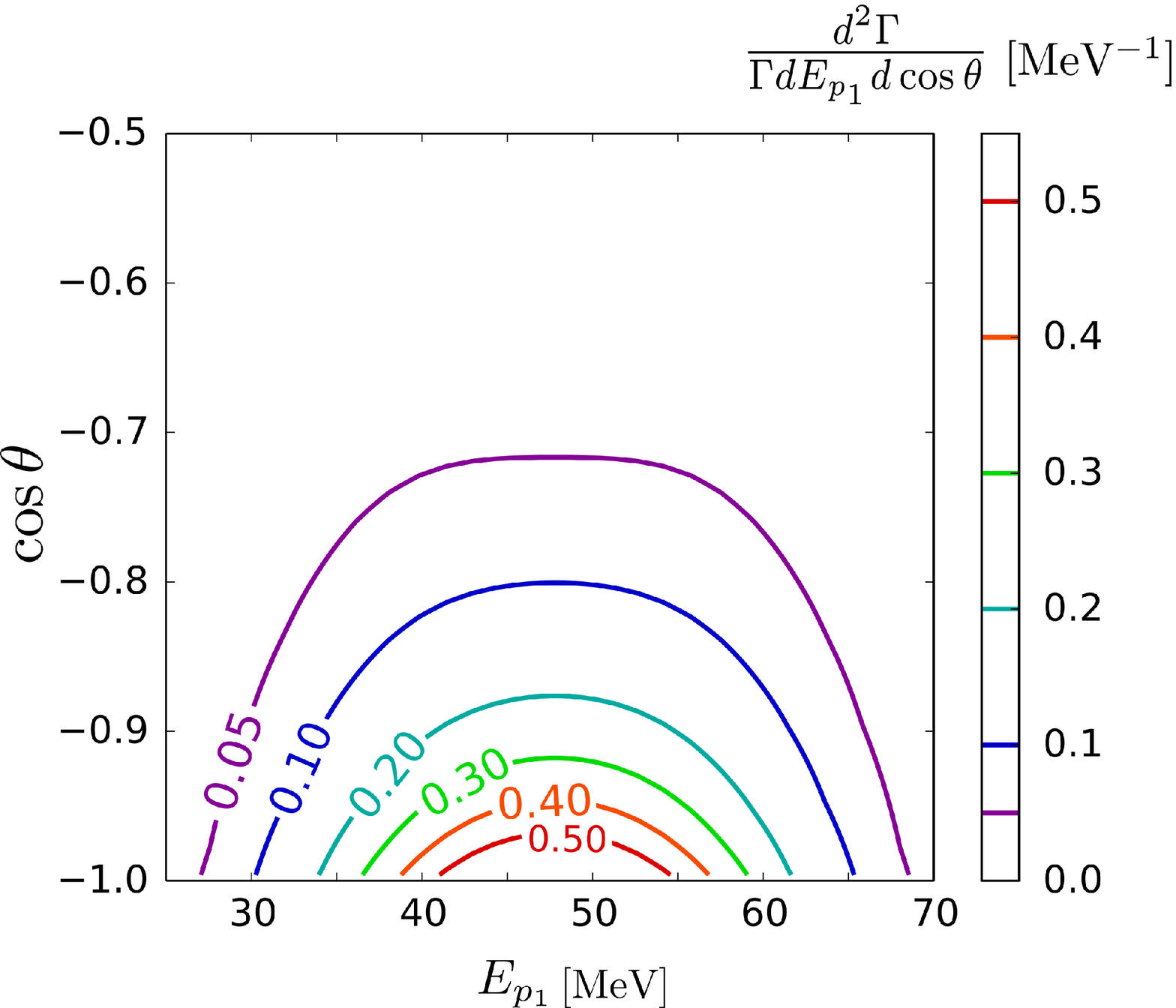}
		\caption{The double differential rate in model (iii) for $^{208}$Pb.}
		\label{fig:ene_ang_distribution}      
      \end{minipage}%
      \hspace{3mm}
      \begin{minipage}{0.49\hsize}
		\centering
		\includegraphics[width=70mm]{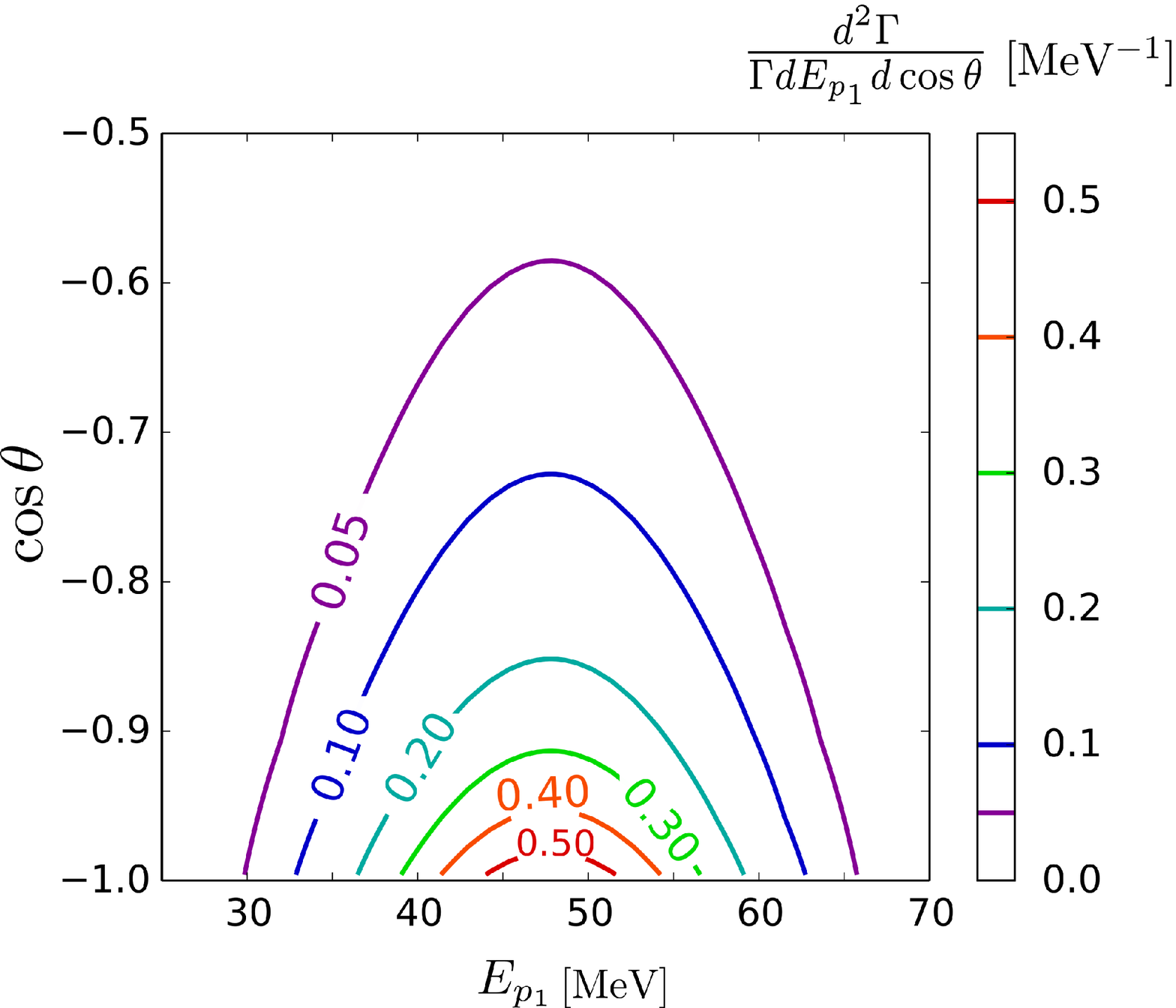}
		\caption{The double differential rate in model (i) for $^{208}$Pb. This figure is printed in Ref.~\cite{Uesaka2016}.}
		\label{fig:distribution_g1}
      \end{minipage}%
    \end{tabular}
\end{figure}

The energy and angular distributions of the emitted electrons also depend on the mechanism of the CLFV interaction. The differential rate of the photonic interaction (model (iii)) and
the contact interaction (model (i)) are shown in Fig.~\ref{fig:ene_ang_distribution} and Fig.~\ref{fig:distribution_g1},
respectively. The tail distributions of backward electrons for the contact interaction are more frequent than for the photonic interaction.
The difference between the model (i) and (ii) appears only when the two electrons are ejected in the same direction
($\cos\theta \sim 1$), where the Pauli principle is most effective, as discussed in \cite{Uesaka2016}.
The distribution of the emitted electrons and the $Z$ dependence of the rate would be useful to identify the mechanism of the CLFV interactions contributing to \muetoee.

\section{Summary \label{sec:Conclusion}}

We have analyzed  the $\mu^-e^-\rightarrow e^-e^-$ CLFV process in muonic atoms.
Together with our previous analysis \cite{Uesaka2016} for the contact interaction
and the present work for the photonic interaction, we find that
the relativistic treatment of the emitted electrons and bound leptons
is essentially important for their qualitative understanding
the rate, in particular the atomic number $Z$ dependence of the rate and the angular and energy
distribution of electrons.
The $Z$ dependence of the \muetoee rate and the distributions of emitted electrons
would be useful to distinguish between the photonic and the four Fermi contact CLFV interactions.
So far one cannot distinguish the $g_1$ term from the $g_2$ term by using these observables.
Therefore the chiral structure of the CLFV interaction should be explored and it would be discussed in our future works. 

\begin{acknowledgments}

This work was supported by the JSPS KAKENHI Grant No. 25105009 (J.S.),
No. 25105010 and 16K053354 (T.S.), No. 25000004 (Y.K.), No. 16K05325 and 16K17693 (M.Y.).
We thank Dr. A. Sato for his fruitful discussions.

\end{acknowledgments}

\appendix
\section{Lepton wave functions}

  The lepton wave functions
  used in our works are given for its completeness\cite{Uesaka2016}.
  The scattering state of an electron
with its momentum $p$ and its $z$-component of spin $s$
for the incoming boundary condition can be expressed as
\begin{align}
\psi^{e(-)}_{\bm{p},s}(\bm{r})=\sum_{\kappa,\nu,m}4\pi i^{l_\kappa}(l_\kappa,m,1/2,s|j_\kappa,\nu)Y_{l_\kappa,m}^*(\hat{p})e^{-i\delta_\kappa}\psi^\kappa_{p,\nu}(\bm{r}),
\end{align}
where $\delta_\kappa$ is a phase shift for the partial wave $\kappa$.
The wave function $\psi^\kappa_{p,\nu}(\bm{r})$ is represented
by the radial wave function $g_p^\kappa(r)$, $f_p^\kappa(r)$ and
the angular-spin wave function $\chi_\kappa$
\cite{rose1961,rose1995elementary} as follows:
\begin{align}
\psi^\kappa_{p,\nu}(\bm{r})=
\begin{pmatrix}
g_p^\kappa(r)\chi_\kappa^\nu(\hat{r}) \\
if_p^\kappa(r)\chi_{-\kappa}^\nu(\hat{r})
\end{pmatrix}.
\label{eq:scattering}
\end{align}
Furthermore, the wave function of a bound lepton $l=\mu,e$ is given as
\begin{align}
\psi^l_{\alpha,s}(\bm{r})=
\begin{pmatrix}
g_{n,l}^\kappa(r)\chi_{\kappa}^s(\hat{r}) \\
if_{n,l}^\kappa(r)\chi_{-\kappa}^s(\hat{r})
\end{pmatrix},
\label{eq:bound}
\end{align}
where $s$ is a $z$-component of spin of the bound state.

\end{document}